\documentclass[aps,prl,reprint,superscriptaddress,nopacs]{revtex4-1}
\usepackage{libertine} 
\usepackage{siunitx}
\usepackage{caption}
\usepackage{graphicx}
\usepackage{dcolumn}
\usepackage{bm}

\begin{document}

\preprint{}

\title{Widefield imaging of superconductor vortices with electron spins in diamond}

\author{Yechezkel Schlussel}
\affiliation{
 Department of Physics, Ben-Gurion University of the Negev, Be'er Sheva 84105, Israel
}%
\author{Till Lenz}
\affiliation{
 Johannes Gutenberg-Universit\"at Mainz, 55128 Mainz, Germany
}%
\affiliation{
Helmholtz Institut Mainz, 55099 Mainz, Germany
}%
\author{Dominik Rohner}
\affiliation{
 Department of Physics, University of Basel, Basel CH-4056, Switzerland
}%
\author{Yaniv Bar-Haim}
\affiliation{
 Department of Physics, Ben-Gurion University of the Negev, Be'er Sheva 84105, Israel
}%
\author{Lykourgos Bougas}
\affiliation{
 Johannes Gutenberg-Universit\"at Mainz, 55128 Mainz, Germany
}%
\author{David Groswasser}
\affiliation{
 Department of Physics, Ben-Gurion University of the Negev, Be'er Sheva 84105, Israel
}%
\author{Michael Kieschnick}
\affiliation{
 Faculty of Physics and Earth Sciences, Felix Bloch Institute for Solid State Physics, Leipzig University, 04103 Leipzig, Germany
}%
\author{Evgeny Rozenberg}
\affiliation{
 Department of Physics, Ben-Gurion University of the Negev, Be'er Sheva 84105, Israel
}%
\author{Lucas Thiel}
\affiliation{
 Department of Physics, University of Basel, Basel CH-4056, Switzerland
}%
\author{Amir Waxman}
\affiliation{
 Department of Physics, Ben-Gurion University of the Negev, Be'er Sheva 84105, Israel
}%
\author{Jan Meijer}
\affiliation{
 Faculty of Physics and Earth Sciences, Felix Bloch Institute for Solid State Physics, Leipzig University, 04103 Leipzig, Germany
}%
\author{Patrick Maletinsky}
\affiliation{
 Department of Physics, University of Basel, Basel CH-4056, Switzerland
}%
\author{Dmitry Budker}
\affiliation{
 Johannes Gutenberg-Universit\"at Mainz, 55128 Mainz, Germany
}%
\affiliation{
Helmholtz Institut Mainz, 55099 Mainz, Germany
}%
\affiliation{
 Department of Physics, University of California, Berkeley, California 94720-7300, USA
}%
\author{Ron Folman}
 \email{Electronic adress: folman@bgu.ac.il}

\affiliation{
 Department of Physics, Ben-Gurion University of the Negev, Be'er Sheva 84105, Israel
}%

\date{\today}

\begin{abstract}
Understanding the mechanisms behind high-$T_{c}$ Type-II superconductors (SC) is still an open task in condensed matter physics. One way to gain further insight into the microscopic mechanisms leading to superconductivity is to study the magnetic properties of the SC in detail, for example by studying the properties of vortices and their dynamics. In this work we describe a new method of wide-field imaging magnetometry using nitrogen-vacancy (NV) centers in diamond to image vortices in an yttrium barium copper oxide (YBCO) thin film. We demonstrate quantitative determination of the magnetic field strength of the vortex stray field, the observation of vortex patterns for different cooling fields and direct observation of vortex pinning in our disordered YBCO film. This method opens prospects for imaging of the magnetic-stray fields of vortices at frequencies from DC to several megahertz within a wide range of temperatures which allows for the study of both high-$T_{C}$ and low-$T_{C}$ SCs. The wide temperature range allowed by NV center magnetometry also makes our approach applicable for the study of phenomena like island superconductivity at elevated temperatures (e.g. in metal nano-clusters\cite{bouchard2011}). 

\end{abstract}

\pacs{Valid PACS appear here}
\maketitle


\section{\label{sec:level1}}

Studying the physics of vortices in Type-II superconductors (SC) is a key challenge in the field of condensed matter physics\cite{Zeldovpinning2015,vortexdynamicszeldov2017,Zeldovscanningsquid2016,Thiel2016,Loudon2015} which in the past was addressed  by various methods. For example, scanning superconducting quantum-interference device (SQUID)\cite{Zeldovpinning2015,Kalisky2010}, scanning Hall-probe magnetometry\cite{Zeldovscanninghallprobe2012,Zeldovscanninghallprobe2006}, Bitter decoration \cite{Grigorieva1994,Rablen2011}, transmission electron microscopy (TEM)\cite{Xia2007,Loudon2015} and magnetooptical imaging\cite{GoaMoke,vorteximagingtechniques1999} were used to image vortices in Type-II superconductors. Recently a new method was proposed\cite{bouchard2011} and realized\cite{Thiel2016}, which uses a single nitrogen-vacancy (NV) center in a scanning diamond tip to study such vortices. Using NV centers enables quantitative noninvasive studies of the vortices over a wide temperature range, which is not achievable by the before mentioned methods.
In this work ensembles of NV centers in diamond are used for an alternative approach of imaging vortices. Utilizing ensembles has the important advantage of not requiring a physical scan of the sample (or the sensor), thereby opening the door for fast imaging up to video frame rates\cite{Horsley2018}. Here, we use this technique in a wide-field microscope to image vortices in an  yttrium barium copper oxide (YBCO) thin film.

NV centers in diamond are point defects in the diamond lattice and have a magnetic field dependent energy-level splitting, which can be driven by resonant microwaves and read out optically by excitation with green light and collection of red NV fluorescence\cite{Rondin2014}.  Magnetometry with NV centers is a widely used technique because of its combination of sensitivity and high spatial resolution. It is applicable in a wide temperature range from cryogenic to ambient temperatures\cite{TempBudker,Thiel2016,Acosta2010a}; in addition, the method enables the determination of both the absolute value and the direction of the magnetic field\cite{Jensen2017,Chatzidrosos2017,Rondin2014}.

\section{\label{sec:level1} Experimental setup}
The experimental setup used in this work incorporates a wide-field imaging microscope (see Fig.\ref{setupodmra}) with the imaging target placed in a continuous-flow helium cryostat. The superconducting sample, a thin YBCO film (thickness\,$\approx250$\,nm), was placed onto the cold finger inside the cryostat and a diamond micro-slab (thickness\,$\approx2\,\SI{}{\micro\meter}$) with \{100\} surfaces was placed flat on top of the superconductor.
The proper engineering of the SC-diamond interface and in particular the resulting NV-SC distance is a crucial enabling factor for this work.
Commonly used diamond plates of dimensions on the order of $3\times3\times0.5\,$mm do not allow to detect individual vortices. Due to the distance between the diamond plate and the SC it was only possible to measure the averaged homogeneous field penetrating the SC\cite{PRBwaxman}.
To overcome this issue, thin diamond plates with dimensions of $20\,\SI{}{\micro\meter}\times10\,\SI{}{\micro\meter}\times2\,\SI{}{\micro\meter}$, with a near-surface NV-center rich layer of $\approx 70$\,nm thickness were fabricated and placed onto the SC (see methods). Microwave fields were applied by a wire (see Fig.\ref{setupodmr}a, inset), which was mounted on the homebuilt cryostat window to keep the heat transfer to the SC as low as possible, while still keeping the wire close to the diamond. The fluorescence was detected using a CCD camera with a pixel size of $16\,\SI{}{\micro\meter}\times16\,\SI{}{\micro\meter}$.

\section*{Results}
\subsection{Optically detected magnetic resonance (ODMR) images}
To generate vortices, the sample was field-cooled in a magnetic field of $\approx0.18$\,mT, applied perpendicular to the surface of the YBCO film. ODMR spectra were taken for every pixel of the camera, each of which corresponding to an area of $0.16\,\SI{}{\micro\meter}\times0.16\,\SI{}{\micro\meter}$ on the sample. The ODMR spectra were obtained by comparing NV fluorescence rates with and without applied microwaves while scanning the MW frequency from 2.84\,GHz to 2.91\,GHz.
The fluorescence contrast was calculated as 
$
C=(N_{\textrm{on}}-N_{\textrm{off}})/(N_{\textrm{on}}+N_{\textrm{off}}),
$
where $N_{\textrm{on(off)}}$ is the photon detection rate while the applied microwave field was on (off). The NV center ground state is a spin-1 system, leading to three states with m$_{s}=-1,\,0$ and $+1$. In the absence of symmetry breaking fields, the m$_{s}=\pm1$ states are degenerate, leading to the same transition frequency from the m$_{s}=0$ state to the m$_{s}=\pm1$ states of $2.87$\,GHz at room temperature ($2.88$\,GHz at cryogenic temperatures). The transition frequencies get split by a magnetic field along the NV axis where the splitting is given by $\Delta\nu_{\pm1}=2\cdot \gamma_{NV}\cdot B_{NV}$, with $\gamma_{NV}=28$\,MHz/mT and $B_{NV}$ the magnetic field along the NV axis. For our diamond samples and at temperatures slightly above the SC phase transition, even at zero field, a splitting ($\approx6$\,MHz) in the ODMR spectra occurred due to strain in the diamond lattice\cite{Rondin2014}. This strain mixes the m$_{s}=\pm1$ states, so that m$_{s}$ are not good quantum numbers anymore and results in a nonlinear relation of $\Delta\nu_{\pm1} \left(B_{NV}\right)=2\cdot\sqrt[]{\left(\gamma_{NV} B_{NV}\right)^2+E^2}$, with E being the strain splitting for low magnetic fields\cite{Doherty2012}. For ensembles of NV centers there are in general four different NV axes with corresponding different field projections that need to be considered. For the \{100\} diamonds used in this work the field projection for a field perpendicular to the surface is the same for all NV orientations.

\begin{figure}
\includegraphics[width=0.5\textwidth]{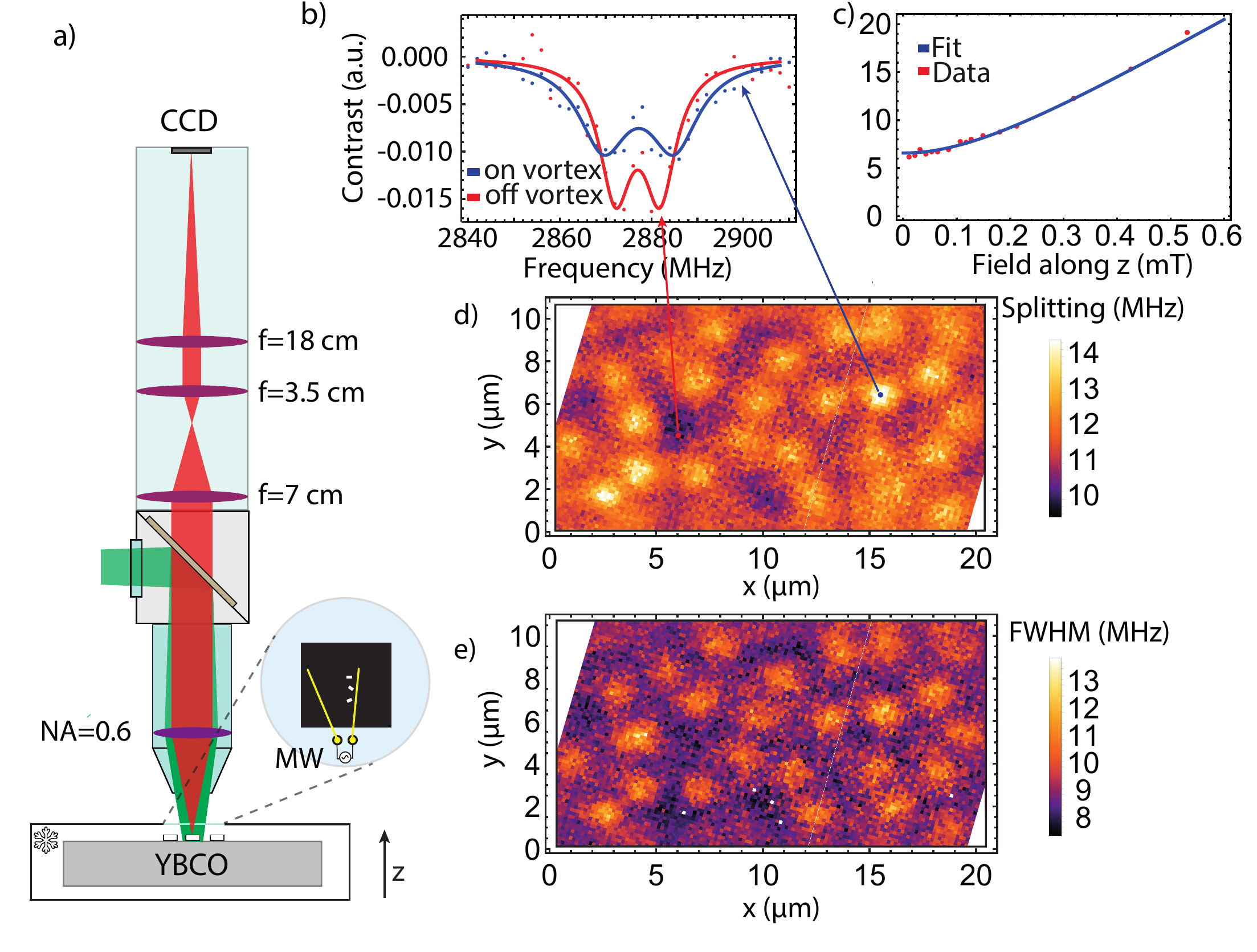}
\caption{a) Schematic of the experimental setup, b) ODMR spectrum for a pixel on top of the vortex and for a pixel outside of the vortex stray field which is $\approx0.38$\,mT in the center of the vortex, c) calibration of the magnetic field in the nonlinear regime at low fields, d) map of the splitting between the two ODMR peaks and e) map of the width of the ODMR peaks. Vortices were identified by an increase in the splitting between the ODMR peaks which is attributed to the magnetic stray field the vortices. In addition, the vortices' stray field leads to a broadening of the peaks, which is generated by field inhomogeneities over the detection volume. The acquisition time for the images d),e) was $\approx11\,$min }

\label{setupodmr}
\end{figure}

To determine the ODMR splitting, the recorded data were fitted to two Lorentzians (Fig.\ref{setupodmr}b) for each pixel, and color maps of the peak width and splitting between the peaks were created (Fig. \ref{setupodmr}d, e). The splitting of the two peaks differed by $\approx4$\,MHz depending on the point of observation with respect to the vortex center. Additionally, the width of the peaks also changes by an amount of 4\,MHz, which is attributed to magnetic field gradients averaged over the optical resolution of $\approx{500}$\,nm and to vibrations which were of similar magnitude as the optical resolution. To illustrate these effects more explicitly, Fig.\ref{setupodmr}b shows two ODMR spectra, one recorded on a vortex and one far from a vortex. 
Even though the change in splitting is on the same order as the splitting due to strain at zero field (see below), it was possible to determine the magnetic field above the vortices. To do this, we calibrated the splitting between the two peaks as a function of magnetic field by applying a known, out of plane magnetic field to the diamond and the data was fitted (see Fig.\ref{setupodmr}c) using the relation mentioned above. The angle between the NV centers and the applied field was considered by using $B_{NV}=B_{z}/\sqrt[]{3}$ and the strain splitting $E$ was determined to be $3.29\left(5\right)\,$MHz. 

Using this calibration the maximum field perpendicular to the SC above the vortices was determined to be between $\approx0.36$\,mT and $\approx0.4$\,mT.
Following the model used in Ref.\cite{simulationpatrick} and considering a bulk London penetration depth $\lambda_{L}=250\,$nm\cite{Thiel2016} these fields correspond to an average distance between the YBCO and the NV layer of $\approx 550$\,nm.
This is larger than expected from the roughnesses of the SC ($\approx100$\,nm) and the diamond plate ($\approx3$\,nm) and the depth of the diamond layer ($d_{NV}\approx15-85$\,nm) and probably originates from dust particles on the surface of the SC and outliers in the roughness of the SC.
Decreasing the distance between the SC and the diamond in the future would shrink the observed size of the vortices and allow us to extract more precise information on the magnetic field.

\subsection{Single-frequency images}
\begin{figure}
\includegraphics[width=0.5\textwidth]{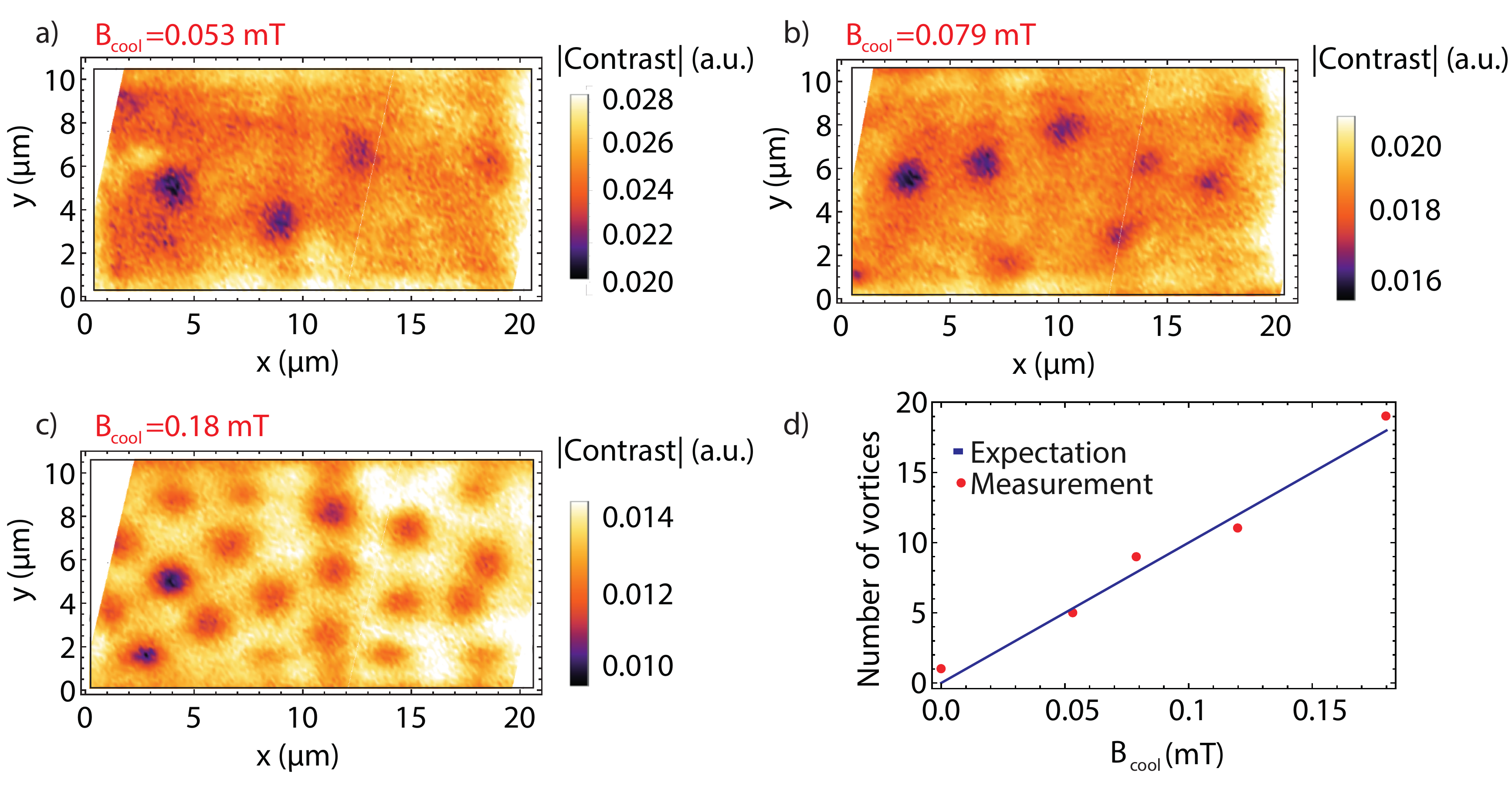}
\caption{a)-c) Single-frequency images (acquisition time $\approx80$\,s) of vortices obtained using microwaves resonant with the NV spin transition at zero magnetic field ($2.876$\,GHz). Images were taken for a magnetic field ($B_{\textrm{cool}}$) present during the cooling through the metal-superconductor phase transition with $B_{\textrm{cool}}=0.053,\,0.079$ and $0.18$\,G. An increase of the number of vortices can be observed when  the field is increased. d) The observed number of vortices as well as the expected number of vortices in the given field of view of $200\,
\SI{}{\micro\meter}^{2}$ for different fields.}
\label{different fields}
\end{figure}

To reduce the image acquisition time, instead of taking whole ODMR spectra for every pixel, images of vortices were obtained by determining the contrast for every pixel for only a single MW frequency (e.g. $2.876$\,GHz) applied to drive the NV spins. The magnetic field of the vortex leads to a splitting of this resonance and therefore to an increased fluorescence at the position of the vortices in the presence of the MW driving field (see Fig.\ref{different fields}). These images were taken for different cooling fields, ranging from $B_{\textrm{cool}}\approx0\,$mT to $0.18$\,mT applied along the direction normal to the SC film. As expected, the different cooling fields led to different vortex densities in the SC. The expected number of vortices is
$
N=B\cdot A/\Phi_{0},
$
where $\Phi_{0}=h/(2e)=2.07 $\,mT$\cdot\SI{}{\micro\meter}^{2}$ is the magnetic flux quantum, $B$ is the magnetic field and $A$ is the investigated area. 
For the dimensions of our sensitive area, which is the area of the diamond micro-slab of $200\,\SI{}{\micro\meter}^{2}$, this led to $N/B\approx100$\,mT$^{-1}$. Figure\ref{different fields}d compares the measured number of vortices with the theoretical expectation and shows good agreement of the two.

\subsection{Pinning}
\begin{figure}
\includegraphics[width=0.5\textwidth]{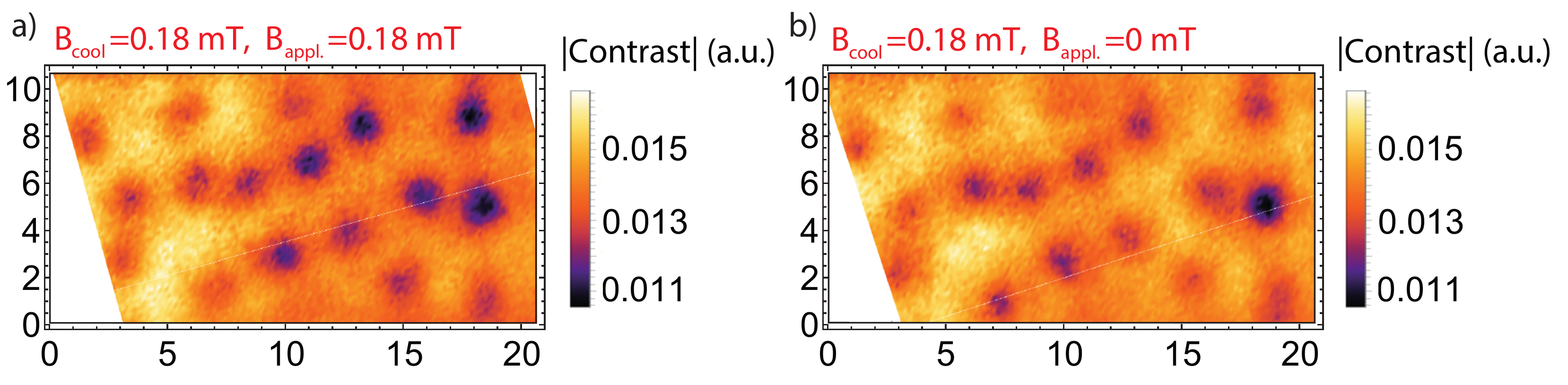}
\caption{a) A single-frequency map (acquisition time $\approx80$\,s) with applied magnetic field and b) A map after the magnetic field was turned off after reaching the SC phase. Both pictures show the same vortex pattern due to strong pinning in the SC sample.}
\label{Pinning}
\end{figure}

Another important phenomenon in Type-II SC is pinning\cite{vortexbible,Zeldovpinning2015}, where magnetic flux (i.e. vortices) is trapped within the SC due to disorder, even when the external field is turned off. This effect is associated with defects which suppress superconductivity and therefore pin the vortices in their positions.  
To study pinning, a single-frequency image of the vortex distribution, with and without applied magnetic field after the cooldown, were taken (Fig.\ref{Pinning}). The same vortices can be identified in both images, which means that pinning is strong enough to trap all vortices which were previously created, extending the results of an earlier study using ensembles of NV centers\cite{PRBwaxman} where individual vortices could not be resolved.

\section*{Conclusion}
A novel technique was presented enabling wide-field imaging of vortices in SCs using NV centers in diamond. By using microfabricated diamond plates, it was possible to reduce the sample-NV distance to less than one $\SI{}{\micro\meter}$, which allowed us to study the stray magnetic field of vortices in a Type-II SC. ODMR spectra were obtained to determine the magnetic-field magnitude from which the distance between SC and diamond was calculated. It was possible to reduce the acquisition time for an entire image using a single frequency.
Single-frequency images were taken for different cooling fields and the obtained number of vortices was compared to the theoretical expectations. In addition, pinning of the vortices in the SC was observed. To improve this technique, one could, for example, use a diamond with less strain to enhance the sensitivity at low magnetic fields or use a \{111\} diamond with preferential NV-center orientation\cite{Lesik2014}, leading to an increase in contrast. In addition, the use of a diamond with \{111\} surfaces would align the NV axis of one of the orientations with the magnetic field of the vortices and therefore increase the resulting splitting in the ODMR spectrum. 
Future experiments will focus on more detailed studies of the vortices and their dynamics, such as vortex oscillations in their pinning potential. Using the presented method such oscillations can be sensed from DC up to several MHz\cite{Acosta2010a,Shin2012}. The implementation of pulsed measurement schemes would in addition enhance the sensitivity to AC fields\cite{Aslam2017,Rondin2014}. Moreover, this technique can be applied as a universal tool to precisely measure magnetic structures of thin films and/or surfaces. 
\section{Methods}

\subsection{Sample preparation}
The diamond plates were fabricated using e-beam lithography and plasma etching\cite{fabricationMaletinsky} of an ``optical grade'' Element 6 CVD diamond with \{100\} surfaces and an initial nitrogen concentration below 1\,ppm. The NV-rich layer close to the surface of the diamond was created using ion implantation of nitrogen ions with energies of 10, 35 and 50\,keV. The diamond was subsequently annealed at 800$^{\circ}$C for 10\,h and at  1200$^{\circ}$C for 2\,h to form the NV centers. Assuming a conversion efficiency of $\approx5\%$\cite{pezzagna2010}, this results in an NV-center density of $\approx3.7$\,ppm within the implanted layer. The plates were broken out of an array\cite{Riedel2014} using a micromanipulator and then placed on the SC. In this procedure, we couldn't control whether the implanted side was facing towards the diamond. To increase the probability of finding a diamond positioned with the correct orientation, several plates were placed on top of the SC.
For a superconductor, a commercially available YBCO thin film from Ceraco consisting of 250\,nm of YBCO grown on a sapphire wafer with a 40\,nm buffer of CeO$_{2}$ was used. This led to an orientation of the YBCO c-axis perpendicular to the surface of the thin film. The critical temperature of the SC sample was $\approx 87$\,K (according to the company specifications).
In addition we were able to transfer the diamond plates to other samples using a sharp needle. This was due to the fact that electrostatic forces between the needle and the diamond were stronger than the van der Waals forces between the SC and the diamond. The low van der Waals interaction is consistent with the surface roughness and the unexpectedly large distance between the SC and diamond plate.
\subsection{Experimental setup}
The experimental setup consists of a continuous-flow helium cryostat (Janis model ST-500) and a homebuilt widefield fluorescence microscope. The green excitation light was generated by a Gem 532 laser (Laser Quantum) reflected with a dichroic mirror (Semrock FF635-Di01-25x36) and focused onto the sample with an Olympus LUCPLFLN40XRC objective with a  40$\times$ magnification and a NA\,$=0.6$. The fluorescence is collected through the same path, after it traverses the dichroic mirror and a telescope (to increase the magnification by a factor of two) it is focused on the camera (Andor iXon 897 electron multiplying charge-coupled device (EMCCD) camera, pixel size $16\,\SI{}{\micro\meter}\times16\,\SI{}{\micro\meter}$) with a  180-mm tube lens.

\section{ACKNOWLEDGEMENTS}
We thank Louis Bouchard, Andrey Jarmola and Victor M. Acosta for fruitful discussions. This work was supported in part by the DFG DIP project Ref. FO 703/2-1, by Swiss NFS grants No. 142497 and 155845 and by the Swiss Nanoscience Institute. This work is also partially supported by the Israeli Science Foundation.

\bibliography{vortices.bib}
\bibliographystyle {ieeetr} 

\end{document}